\documentclass[runningheads]{llncs}
\usepackage{graphicx}
\usepackage{comment}
\usepackage{booktabs}
\usepackage{multirow}
\usepackage[table,xcdraw]{xcolor}
\usepackage{soul}
\usepackage[backref=page]{hyperref}
\usepackage{subcaption}
\begin{document}
\title{A Novel Hybrid Endoscopic Dataset for Evaluating Machine Learning-based Photometric Image Enhancement Models}
\titlerunning{A Novel Hybrid Endoscopic Dataset for testing ML-based IE methods}
%
\author{Axel García-Vega\inst{1} \and Ricardo Espinosa\inst{2,3} \and  Gilberto Ochoa-Ruiz\inst{1} Thomas Bazin \inst{4} \and Luis Eduardo Falcón-Morales \and \inst{1} \and Dominique Lamarque \inst{4} \and Christian Daul \inst{4}}
\authorrunning{A. García et al.}
%
\institute{Tecnologico de Monterrey, Escuela de Ingenieria y Ciencias \\ Monterrey, Nuevo León, México \and Universidad Panamericana, Facultad de Ingenier\'ia \\ Josemaría Escrivá de Balaguer 101, Aguascalientes, 20290, M\'exico  \and Université de Lorraine and CNRS, CRAN UMR 7039, 2 avenue de la Forêt de Haye, F-54518 Vandœuvre-Lès-Nancy Cedex\and Hôpital Ambroise Paré, 9 av. Charles de Gaulle, F-92100 Boulogne-Billancourt}

\maketitle              
%


\begin{abstract}
Endoscopy is the most widely used medical technique for cancer and polyp detection inside hollow organs. However, images acquired by an endoscope are frequently affected by illumination artefacts due to the enlightenment source orientation. There exist two major issues when the endoscope's light source pose suddenly changes: overexposed and underexposed tissue areas are produced.
These two scenarios can result in misdiagnosis due to the lack of information in the affected zones or hamper the performance of various computer vision methods (e.g., SLAM, structure from motion, optical flow) used during the non invasive examination. The aim of this work is two-fold: i) to introduce a new synthetically generated data-set generated by a generative adversarial techniques and ii) and to explore both shallow based and deep learning-based image-enhancement methods in overexposed and underexposed lighting conditions. Best quantitative results (i.e., metric based results), were obtained by the deep-learnnig -based LMSPEC method, besides a running time around 7.6 fps.\\

\textbf{Data available at:}
https://data.mendeley.com/datasets/3j3tmghw33/1

\end{abstract}
\section{Introduction}
\label{introduction}

\subsection{Medical Context}
\label{context}

Endoscopy is the most effective and common used examination tool to prevent colon cancer by screening for lesions. This technique is used to examine the human colon through a flexible tube called endoscope, while a camera attached at the tip gathers visual information in real-time. Additionally to the camera, a point light provides the lighting source during the surgery.

On the other hand, Minimally Invasive Surgery (MIS) is a type of endoscopic procedure
 that is increasingly becoming a mainstream medical procedure, overtaking traditional surgical operations and treatments. Such endoscopic intervention entails a less traumatic experience and less pain for the patient, quicker recovery after surgery and shortened hospital stays \cite{fu2021future}. 
 
\begin{figure}[!t]
    \centering
    \includegraphics[scale=0.35]{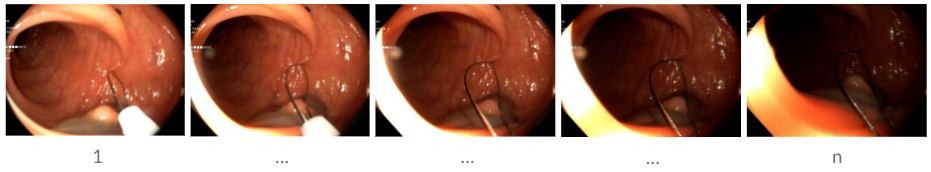}
    \caption{Light condition changes in sequential video frames. Frame $1$ shows a normal condition. Later frames show poor conditions while light points to the closest object.}
    \label{sequential}
\end{figure}

 Both endoscopic examinations and MIS procedures require a great deal of expertise from the surgeon or specialist and demand complicated training and certifications. Despite of this, there is ample evidence in the literature that during the vast majority of endoscopic examinations, a great deal of regions of interest are missed \cite{ma2019real}, which poses a serious problem, as they might contain suspicious regions or other lesions such as polyps. Moreover, such issues make the inspection by the physician a strenuous and time consuming task. In general, these problems stem from how rapidly the lighting conditions in the endoscopic video can change from frame to frame (see Fig. \ref{sequential}). In fact, a highly  non-linear illumination response produces endoscopic frames which are  highly lit in some sections (overexposed) and poorly lit (underexposed) in other areas. 

Additionally to  affecting strongly the efficiency in the lesion detection by the doctor, the highly changing illumination conditions in endoscopic settings also hampers the performance of AI-based tools that are being increasingly developed for Computer-aided Detection (CADe) and Computer-aided Diagnosis (CADx) applications and in MIS and laparoscopy, among other areas of research, as we will discuss next.

\subsection{Motivation for our proposal}
\label{motivation}

In recent years  computer vision (CV) has been playing significant role in endoscopic explorations aided by novel deep learning (DL) techniques, which give computers the visual and temporal learning abilities to understand complex surgical procedures in hollow organs \cite{fu2021future}. The integration of such techniques is currently being investigated also for laparoscopy and MIS applications (i.e., such as instrument tracking, endoscopic view enhancement and suspicious lesion tracking \cite{fu2021future}). Among these applications, 3D surface reconstruction seems to be a promising solution for the poor depth perception which prevents the full-fledged adoption of the above-mentioned techniques in MIS and Robotic-assisted surgery (RAS). 

However, as with other applications of computer vision in endoscopy, proper illumination conditions is an aspect of utmost importance for attaining a high performance in applications such a CADx and such reconstruction methods. In the latter case, the techniques traditionally employed for recovering 3D information, such as Simultaneous Localization and Mapping (SLAM) \cite{chen2018slam}, Structure for Motion (SfM) and Optical Flow (OF) \cite{phan2020optical}, work poorly on endoscopic images. This is due to strong photo-metric variations caused by moving light sources, moist surfaces, occlusions as well as reflections for surgical instruments that provoke underexposed and overexposed video frames \cite{ma2019real}. 

As a matter of fact, 3D reconstruction techniques rest on the constraint of brightness constancy \cite{phan2020optical}. This constraint assumes that the values in intensity of the pixels remain constant or with a few variations, which usually holds in applications in natural imagery (i.e., photography) \cite{afifi2021learning} or autonomous driving \cite{elgendy2020deep}. However, in endoscopic procedures, as Figure \ref{sequential} shows, the brightness constancy assumption does not hold due to the numerous illuminations variations when the camera and light move through the organs. 

Therefore, the use of image enhancement techniques for pre-processing such images is a mandatory  step to carry out a robust 3D reconstruction \cite{shao2022self} and to develop reliable and robust CADe/CADx tools.

\subsection{Contributions and organization of the article}
\label{contributions}

Over the decades, multiple image enhancement techniques have been explored for illumination adaptation, exposure correction and high-dynamic-range tone mapping in several applications, to cite a few. Nevertheless, most of the proposed methods for exposure correction have been designed to deal either with low-light  high-light settings separately, due to the different characteristics of these enhancing tasks in different areas. In enhancing endoscopic images, these methods have presented a general lack of robustness due to the fact that in endoscopy images both problems (under/over exposure) are present simultaneously.

Additionally, as we will discuss next, most of the machine learning based approaches to IE  require large amounts of paired-data images (i.e., corrupted and non-corrupted ground truth images) for training and testing purposes. However, to the best of our knowledge, such a large database does not exist for testing image enhancement algorithms in endoscopic images.

Therefore, in  order to mitigate the problems related to large variations in illumination in endoscopic interventions, in this paper we present novel synthetic dataset for testing and evaluating endoscopic image enhancement methods. In order to attain this goal, of our proposal consists of the following contributions:

\begin{itemize}
    \item 1. We introduce a novel dataset containing synthetically generated over and under exposure frames, in tandem with their respective ground truth images. We believe that this dataset can serve as a cornerstone for testing well known IE methods and provide a baseline for future developments in ML for endoscopy and in DL-based image enhancement methods.
    \item 2. We perform a thorough evaluation of traditional and DL-based image enhancement (IE) methods in order to highlight the importance of the ground truth data and the scarsity of bidirectional (under/over exposure) methods. 
\end{itemize}




The rest of this paper is organized as follows. In Section \ref{data_methods} we present the datasets from which the raw data was obtained, besides the methodology utilized for creating the synthetic dataset and both traditional and DL-based methods for enhancing the synthetic data. In Section \ref{E&R} we provide a thorough description of the data preparation and training setup for our experiments. Also, we present a brief context about metrics implemented and their respective results. Finally, in Section \ref{Conclusion_FW}, we give our conclusion and an ideas for future work.

\section{State of the art}
\label{State-of-the-art}

Mainly, image enhancement consists of two main domains i) spatial domain that involves direct manipulation of pixels, and ii) frequency domain that involves Fourier transformation of the input image \cite{LOL}. IE methods improve the perception of the image using spatial domain, frequency domain and a combination of both methods in order to output a modified image on its contrast, hue, or brightness.

These methods have been approached  using either traditional or deep learning techniques. Some examples of the traditional methods group on the spatial domain are Histogram Equalization (HE) \cite{wang2008flattest}, Dehazing-based methods \cite{savelli2017illumination} and curve adjustment approaches \cite{huang2012efficient}. Another category of image enhancement approaches are those based on Retinex theory,  which decompose the images in reflection and illumination components \cite{LOL}; examples of these methods are Single Scale Retinex (SSR) \cite{rahman1996multi} and multi-scale Retinex (MSR) \cite{rahman1996multi}. These methods suffer from a limitation in model capacity for the decomposition and it is difficult to implement successfully in endoscopy scenarios due to the challenging lighting changes. 

Recently, deep learning based methods have been explored for achieving more accurate and real-time image enhancement procedures with successful results. However, most the works in the literature have been oriented to the low-light image enhancement such as deep-learning networks based on the Retinex assumption, which combines Retinex assumption with the Convolutional Neural Networks (CNN) \cite{LOL}.

Furthermore, although these methods have shown great promise in natural images, many of them cannot be directly applied to endoscopic  images or videos. First of all, most of the existing methods can either enhance either under-or over-exposed images but not both. For instance, Deep UPE \cite{wang2019underexposed}, or the work presented in \cite{zhang2021beyond,zhang2021lighting} are examples of methods that can be applied successfully in under-exposed natural images, whereas just a few methods have been explored to correct both under-exposed and over-exposed images. Recently, Afifi et al. \cite{afifi2021learning} proposed a novel deep learning based technique to enhance images in both directions under- and over-exposure in a supervised manner.

Herein,  we benchmark this technique against traditional methods in order to assess our real-synthetic paired dataset and highlight the need for reliable and standard datasets for future DL-based IE methods in endoscopic procedures. Nonetheless, another requirement for developing usable IE methods in endoscopy is related their inference time or latency. Most of the works found in the literature take several seconds to enhance a single image, hampering  their use in real-time image processing applications such as video processing of underwater \cite{li2019underwater}, X-ray imaging \cite{naidumedical} and restoration in video endoscopy \cite{fu2021future} . Endoscopy is an application area which requires dynamic and real-time image enhancement, which additionally needs to be accurate in the sense of not introducing any artifacts and we believe that the work presented in \cite{afifi2021learning} can serve a launching pad to satisfy these requirements.

Figure \ref{fig:overunder} shows some examples of endoscopic video frames with some of these examples. These dramatic changes impede the development of robust computer vision methods for CADe and CADx among other tasks such as MIS or CIS (computer integrated surgery). Besides, it has been demonstrated that image enhancement pre-processing techniques can significantly boost the performance of
3D reconstruction pipelines using endoscopic images \cite{zhang2021lighting}. 

As detecting artifacts such as under and over exposed frames is so important for a variety of applications, and in other to improve the applicability of CADe and CADx application endoscopy, several datasets and challenges have have been proposed \cite{ali2020translational,ali_ghatwary}. The focus of these challenges has been to foster the development of real-time methods for detecting artefacts such as bubbles, instruments, blood or even some lesions, where under and overexposure play a major role, as they are typically discarded in many automated procedures. However, many computer vision algorithms such as SfM or SLAM require as many frames as possible for maximizing the quality of the obtained results and as such, discarding such frames is not an option and image enhancement is thus necessary. 

An additional problem is that, to the best of our knowledge, datasets containing both pairs of  ground truth (clean images) and corrupted images (with exposure errors) does not exist in for endoscopic imaging. This is in stark contrast to the more conventional/mainstream computational photography research field, in which several standard datasets have been proposed and are widely used for testing new algorithms and architectures \cite{LOL,fivek}

Thus, in order to develop and test new image enhancement methods in endoscopy is necessary to develop and test such dataset from the ground up. The contribution of the work presented in herein is thus creation of such a novel real-synthetic paired dataset. For doing so, we leverage three existing endoscopic datasets and a Generative Adversarial Networks (GANs) approach for transferring under and over exposed over clean images. In this manner, we have both a reference image and a corrupted images over which reference-based metrics  (i.e. SSIM) can be applied in order to assess the performance of newly developed enhancement methods. 

\section{Data and Methodology}
\label{data_methods}

\subsection{Data}

\begin{figure}[!t]
     \centering
     \begin{subfigure}[t]{0.32\textwidth}
         \centering
         \includegraphics[width=3.5cm,height=3.5cm]{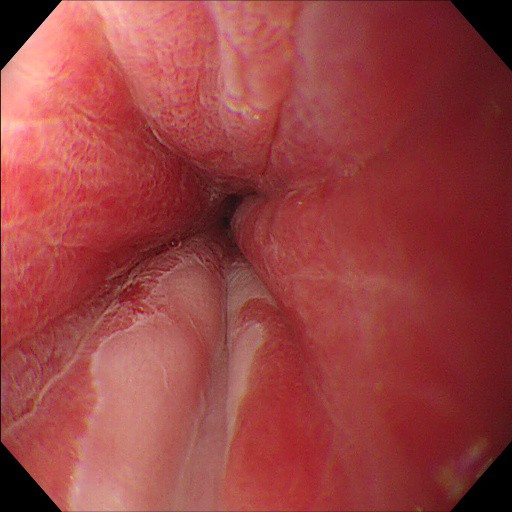}
         \caption{Without exposure error.}
         \label{normal}
     \end{subfigure}
     \begin{subfigure}[t]{0.34\textwidth}
         \centering
         \includegraphics[width=3.5cm,height=3.5cm]{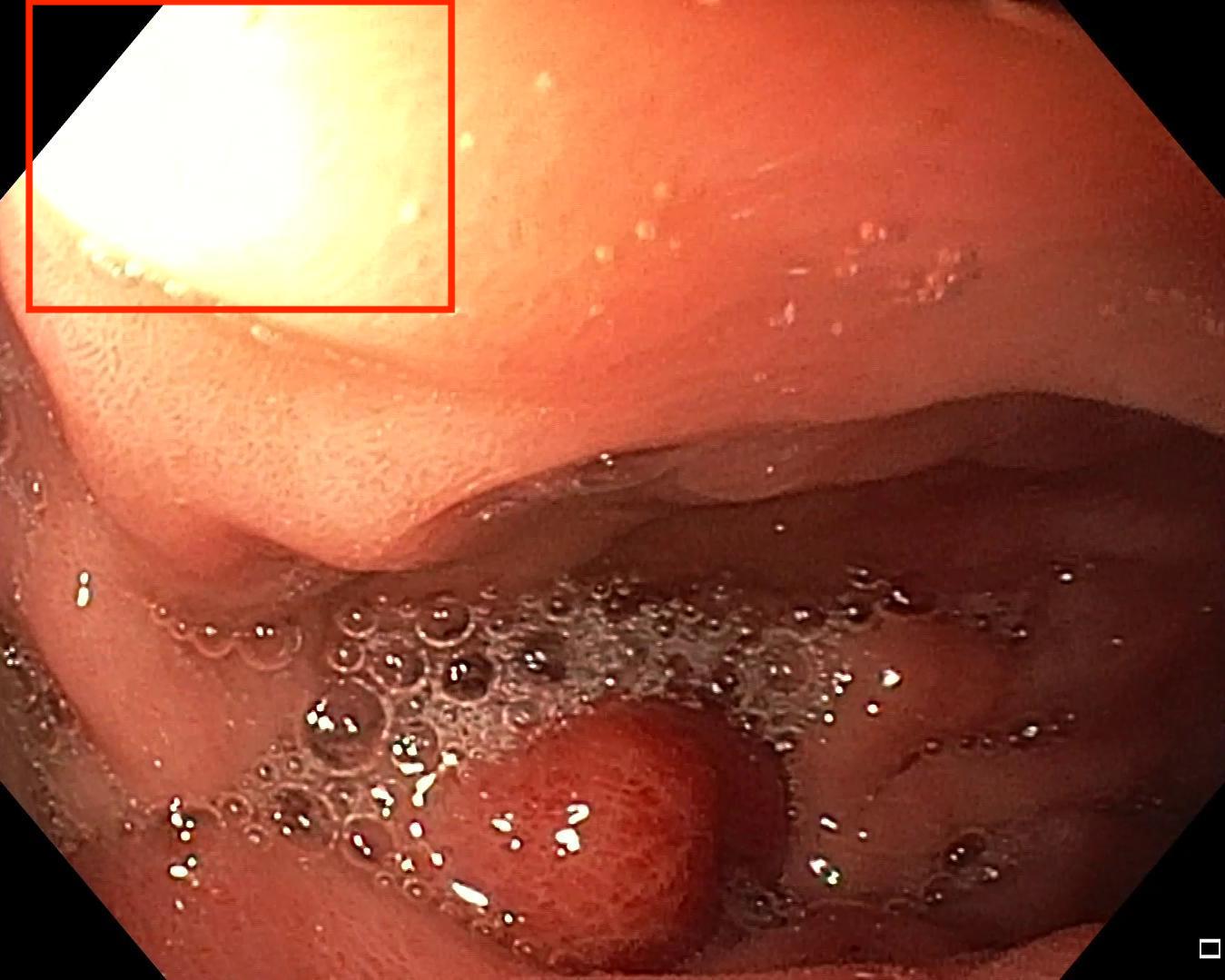}
         \caption{Overexposure.}
         \label{overexp}
     \end{subfigure}
     \hfill
     \begin{subfigure}[t]{0.32\textwidth}
         \centering
         \includegraphics[width=3.5cm,height=3.5cm]{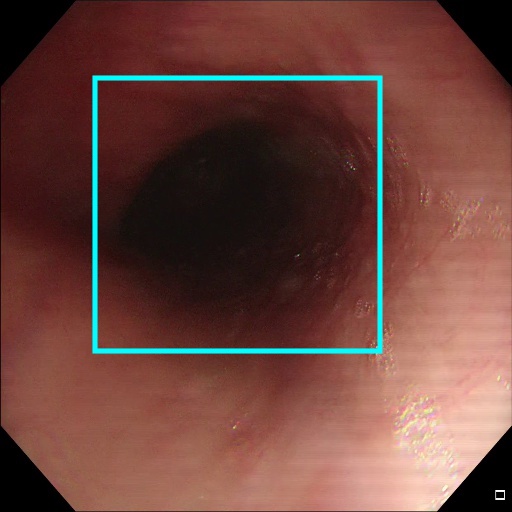}
         \caption{Underexposure.}
         \label{underexp}
     \end{subfigure}
        \caption{Example of endoscopic frames exhibiting the types of artefacts that we are are interested on removing through image enhancement methods }
    \label{fig:overunder}
\end{figure}

As briefly discussed above, the data for our experiments was obtained from three different sources, none of them containing enough information for doing image enhancement with deep learning-based methods, i.e., these are not datasets with paired frames for applying methods where we have access to the corresponding ground truth. The chosen datasets are: the  EAD2020 Challenge dataset \cite{ali2020translational}, EAD 2.0 from EndoCV2022 Challenge dataset \cite{ali_ghatwary} and the HyperKvasir dataset \cite{Borgli2020}. They were selected for our experiments as they contain single frames and also sequential frames from several hollow organs. Figure \ref{fig:overunder} shows examples of a normal frame (without exposure error) and two examples containing exposure problems that were extracted from the datasets.

\subsubsection{EAD2020 Challenge Dataset}
\label{EAD2020}

The Endoscopy Computer Vision Challenge (EndoCV2020) addressed two challenges both focused on finding novel methods for detection and segmentation in the endoscopic videos: Endoscopy Artifact Detection (EAD2020) and Endoscopy Disease Detection (EDD2020) \cite{ali2020translational} challenges. The goal of the EDD2020 sub-challenge was to develop methods for detecting and segmenting visible diseases. For this purpose, a full dataset comprised of images, annotations and masks was created from video frames and annotated by experts from various institutions. However, the nature of the challenge involves only classes related to diseases. On the other hand, EAD2020 challenge comprises a variety of endoscope positions, organs, disease/abnormality and image artefacts.  All frames were annotated by experts with 8 different classes; in particular, two classes are of interest to our research: contrast and saturation artefacts. These classes denote what is well-known in photography as overexposure and underexposure errors but in specific areas on the image. From the 2531 images, only 770 frames were labeled with underexposure (contrast) and 249 with overexposure (saturation). Figure \ref{fig:overunder} shows two examples from the dataset containing these two kinds of artifacts.

\subsubsection{EAD2.0 Dataset}
\label{EAD2.0}

Similarly to EAD2020 Challenge Dataset in Sec. \ref{EAD2020}, EAD2.0 Dataset \cite{ali_ghatwary} consists of multi-center and diverse population sub-datasets with tasks for detection and segmentation but focus on assessing generalizability of algorithms. This dataset contains more sequence/video data and multi-modality data from different centers. Particularly, the dataset consists of 24 sequences with 1106 video frames with multi-instance labels for training. However, unlike the EAD2020, EAD2.0 dataset was not labeled with any class related to underexposure errors. In Section \ref{Data_prep} we will explain briefly the way we filtered out the dataset in order to have images with exposure errors.

\subsubsection{HyperKvasir Datset}
\label{HyperKvasir}

Borgli et. al in \cite{Borgli2020} presented the largest image and video dataset of the gastrointestinal tract available in literature with 110,079 images and 374 videos. The data was collected during real gastro- and colonoscopy examinations. The dataset is split into labeled images (10,662) and unlabeled images (99,417). Since this dataset had no a specific purpose, data was not annotated, and in fact the only labels that it contains are related to the organ type that the video frames comes from. In present work we only take the set of labeled images.\\

\subsection{Methodology}
\label{Methodology}


Given a set of raw endoscopic frames collected from the three datasets previously mentioned, our pipeline aims to i) train an object detector to classify frames with and without exposure errors over unlabeled datasets (see Fig. \ref{OD}), ii) manually filter out non-informative frames, iii) train a GAN for creating synthetic frames with exposure errors (see Fig. \ref{pipeline}), and iv) quantitatively and qualitatively assess our dataset on various types of image enhancement methods:  traditional algorithms based on histogram equalization, models based on Retinex theory and deep neural network architectures.

\subsubsection{Data Preparation}
\label{Data_prep}

\begin{figure}[!t]
     \centering
     \begin{subfigure}[c]{0.7\textwidth}
         \centering
         \includegraphics[scale=0.3]{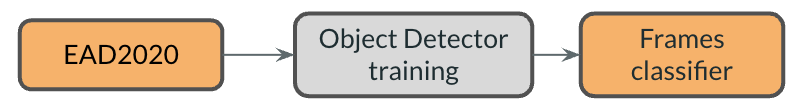}
         \caption{Object detector for classifying frames.}
         \label{OD}
     \end{subfigure}
     \\[0.5cm]
     \begin{subfigure}[c]{0.85\textwidth}
         \centering
         \includegraphics[scale=0.48]{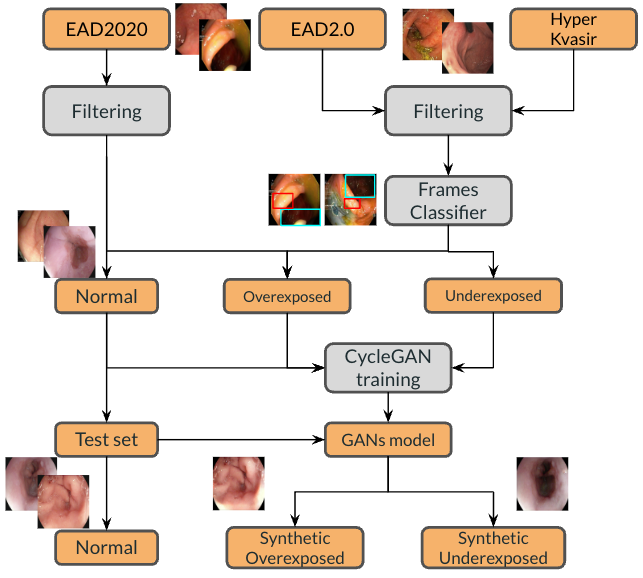}
         \caption{Pipeline for creating our real-synthetic endoscopy dataset for IE.}
         \label{pipeline}
     \end{subfigure}
     \caption{Overall pipeline of the methodology}
    \label{fig:overall_pipeline}
\end{figure}

We discarded non-informative frames. For instance, fully dark, bright or blurry images, were filtered out. After this process, we ended up with 7,064 images, 1049 from EAD2020, 654 from EAD2.0 and 5361 from HyperKvasir.As we mentioned before, the EAD2020 dataset already contains annotations about exposure errors of our concern, hence we trained a YOLOv4 object detector \cite{bochkovskiy2020yolov4} with data from EAD2020 \cite{ali2020translational} and we obtained reliable results on detecting exposure errors. Finally, the model was applied to detect those instances in EAD2.0 \cite{ali_ghatwary} and HyperKvasir \cite{Borgli2020}. Therefore, as shown in Figure \ref{fig:overall_pipeline}, the object detector acts as a frame classifier, since it outputs frames without exposure errors, and with over and under exposure errors.

\subsubsection{Image-to-Image Translation}
\label{Im-to-Im}


Image-to-image translation is an application of GANs which manage to translate from one representation of an image to another, or rather it does style transfer. For instance, a scene may be rendered by a grayscale image, RGB image or edge sketches \cite{elgendy2020deep}.

One of the contributions of the present work is to use image-to-image translation \cite{elgendy2020deep} to take normal endoscopic frames and transfer to style with over- and underexposure.  For implementing this task, we used the CycleGAN architecture \cite{zhu2017unpaired} since the main problem to tackle is the lack of paired data, thus CycleGAN is a promising method for working with our unpaired data. 

\vspace{-4mm}

\subsubsection{Image Enhancement}
\label{Im_Enh}


One of the concerns of this work has been to find image enhancement methods that perform the restoration of both types of exposure errors, since these problems are often found simultaneously in endoscopic examinations. Contrary to other methods in the literature, the model proposed by Afifi et al. \cite{afifi2021learning}is capable of handling both problems in natural images with accurate results and with a very low latency and we will used as a baseline for comparison traditional image processing algorithms as well as for future developments.

The approach proposed by the authors, known as Learning Multi-Scale Photo Exposure Correction (LMSPEC)  extracts random patches of three different sizes (128x128, 256x256 and 512x512) and decomposes each patch in four-level frequencies, i.e., it makes a Laplacian Pyramid (LP) of four levels. Then, each level of the resulting LP is utilized as input of a set of sequential U-Net-like sub-networks. For each patch, an L1 loss-based loss functions is proposed, aimed to store global color information besides detail information. A discriminator network is also for computing an adversarial loss aimed to preserve realism in the corrected patch. In this work, we implemented a quantitative and qualitative benchmark using i) traditional methods: RLBHE \cite{zuo2013range} and FHSABP \cite{wang2008flattest} based on histogram equalization, and LIME \cite{guo2016lime} and DUAL \cite{zhang2019dual} based on Retinex theory, and ii) DL-based method: LMSPEC \cite{afifi2021learning}. 

\vspace{-3mm}

\subsection{Training Setup}
\label{Train_setup}

All our models were trained and the experiments were executed on an NVIDIA DGX-1 system with eight Tesla V100 GPUs.

\vspace{-3mm}

\subsubsection{YOLOv4 setup}
\label{YOLOv4_setup}

Frames annotated by experts provided on the EAD2020 challenge \cite{ali2020translational} allowed us to train YOLOv4 object detector (for classifying frames without exposure errors, under- and over-exposed) with 90\% for training and 10\% for testing. For avoiding over-fitting, data augmentation techniques such as rotations, splits, blurry, and hue change, were applied. The hyper-parameters were set as follows: the number of classes were 2 (overexposure and underexposure), thus the number of filters on each convolutional layer was 30. The training steps was set to 6000 with an initial step decay learning rate of 0.01 and divided by factor of 10 at the 4,800 steps and 5,400 steps. The momentum was set as 0.9 and weight decay as 0.0005.

\vspace{-3mm}

\subsubsection{CycleGANs setup}
\label{CycleGANs_setup}

For the first experiment all frames with exposure errors, i.e., $1,296$ overexposed and $1,289$ underexposed, were taken as adversarial discriminators $D_{Y}$. On the other hand, the $4,478$ normal frames were split up into 65\% for training and 35 \%for testing. The hyper-parameters were set for both experiments as follows: 150 epochs, ADAM optimizer, learning rate of 0.0002, decay $\beta_{1}$ of $0.5$, decay $\beta_{2}$ of 0.999, starting decay from epoch $100$, cycle loss weight $\lambda_{cyc}$ of $10$ and identity loss weight $\lambda_{id}$ of 5.

\subsubsection{LMSPEC setup}
\label{LMSPEC_setup}

 We perform the same split for manage both exposure errors, 70\% for training, 27\% for testing and 3\% for validation. We only perform experiments on patches with dimensions 128x128 and 256x256. The hyper-parameters were setup for both experiments as follows: ADAM optimizer, decay rate $\beta_{1}$ of $0.9$, decay $\beta_{2}$ of 0.999, a learning rate of 0.0001 for the generator and 0.00001 for the discriminator; for patches with dimension 128x128 we used a mini-batch size of 32, 40 epochs and at epoch 20 the learning rate was decayed by factor of 0.5. For patches with dimension 256x256 we used a mini-batch size of 8, 30 epochs and each 10 epochs the learning rate was decayed by the same factor; the adversarial loss was activated only for patches with dimensions 256x256 after 15 epochs.

 \subsection{Metrics}
\label{Metrics}

One of the drawbacks of implementing a framework like ours, which is based in GANs and synthetic datasets, is that there are not other works to compare against our results. Therefore, for evaluating our image enhancement results we make use of well-known of reference-based metrics, thus the need for a paired dataset with ground truth images. For instance, in \cite{welander2018generative} P. Welander et al. performed a perceptual study for evaluating the realism on the resulting synthetic images. Unlike them, herein we performed a perceptual study for evaluating the quality of the generated images. First of all, we use the Structure Similarity Index (SSIM) for filtering out similar frames, whereas the Main Square Error (MSE) on the gray-scale domain was used for filtering out very darkened/brightened frames, or for discarding frames without changes on light conditions after the GAN style transfer. 

On the other hand, for evaluating our image enhancement experiments we implemented a statistical analysis, ground truth dependent evaluation. We used the Mean Squared Error (MSE) to evaluate the pixel-wise average squared errors in the image. We also make use of of the Peak Signal-to-Noise Ratio (PSNR) and of Structural Similarity Index Metric (SSIM) \cite{wang2004image} for evaluating quantitatively the quality of the enhanced images results.

\vspace{-3mm}

\section{Experiments and Results}
\label{E&R}

We implemented three main experiments. i) we created a frames classifier by training and testing YOLOv4 object detector, ii) we created a paired real-synthetic dataset by training and testing CycleGANs, and iii) we utilized our dataset for exposure correction with traditional methods, and training and testing LMSPEC to highlight the importance of a paired data. 

\vspace{-3mm}

\subsection{Results}
\label{Results}

After implementing the CycleGAN model, we obtained 1,564 frames for each type of exposure error. However, some frames were not useful or sufficiently informative and thus we carried out a statistical analyses based on the MSE and SSIM metrics that were used to establish boundaries that separate informative from non-informative frames. For our dataset, the range for the SSIM was between  $0.6\leq SSIM < 0.9$ ; on the other hand, the value of the MSE over grayscale levels boundaries was $100<MSE<1,500$ for overexposure and $100<MSE<750$ for underexposure errors. After this procedure, we finally obtained a dataset with 985 underexposed and 1,231 overexposed frames ground truth and corrupted images.

\addtolength{\tabcolsep}{4pt} 
\begin{table}[!t]
\centering
\caption{\small{Results for full reference-based quality experiments.}}
\label{tab:t2}
\begin{tabular}{cccccccc}
       & \multicolumn{3}{c}{Overexposure}                  & \multicolumn{3}{c}{Underexposure}                 &                 \\
Method & MSE$\downarrow$            & PSNR$\uparrow$            & SSIM$\uparrow$           & MSE$\downarrow$            & PSNR$\uparrow$            & SSIM$\uparrow$           & Inference time  \\ \hline
LIME   & 0.048          & 8.566           & 0.597          & 0.054          & 17.331          & 0.699          & 44.0236         \\
DUAL   & 0.043          & 0.907           & 0.726          & 0.053          & 20.012          & 0.708          & 30.9965         \\
FHSABP & \textbf{0.036} & 16.021          & 0.631          & 0.034          & 18.195          & 0.633          & 0.2953          \\
RLBHE  & 0.051          & 19.425          & 0.746          & 0.055          & 21.053          & 0.723          & 0.2996          \\
LMSPEC & \textbf{0.046} & \textbf{24.061} & \textbf{0.812} & \textbf{0.964} & \textbf{23.863} & \textbf{0.793} & \textbf{0.1316} \\ \hline
\end{tabular}
\end{table}


\begin{figure}[!t]
     \centering
     \begin{subfigure}[c]{\textwidth}
         \centering         \includegraphics[scale=0.22]{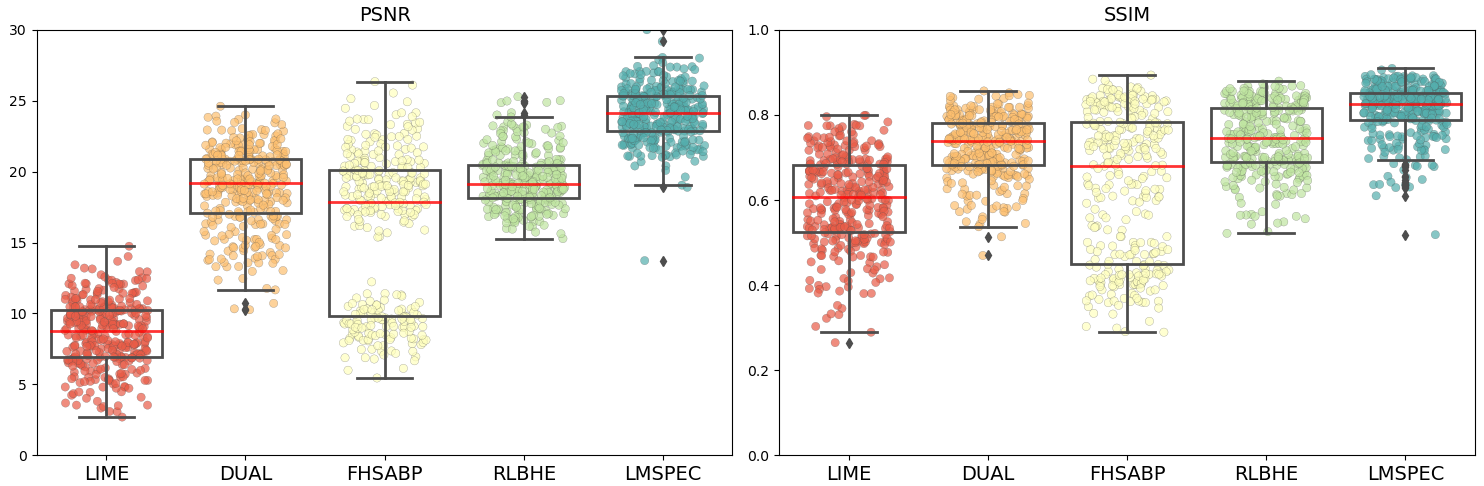}
         \caption{Overexposure.}
         \label{overexp}
     \end{subfigure}
     \hfill
     \begin{subfigure}[c]{\textwidth}
         \centering
         \includegraphics[scale=0.22]{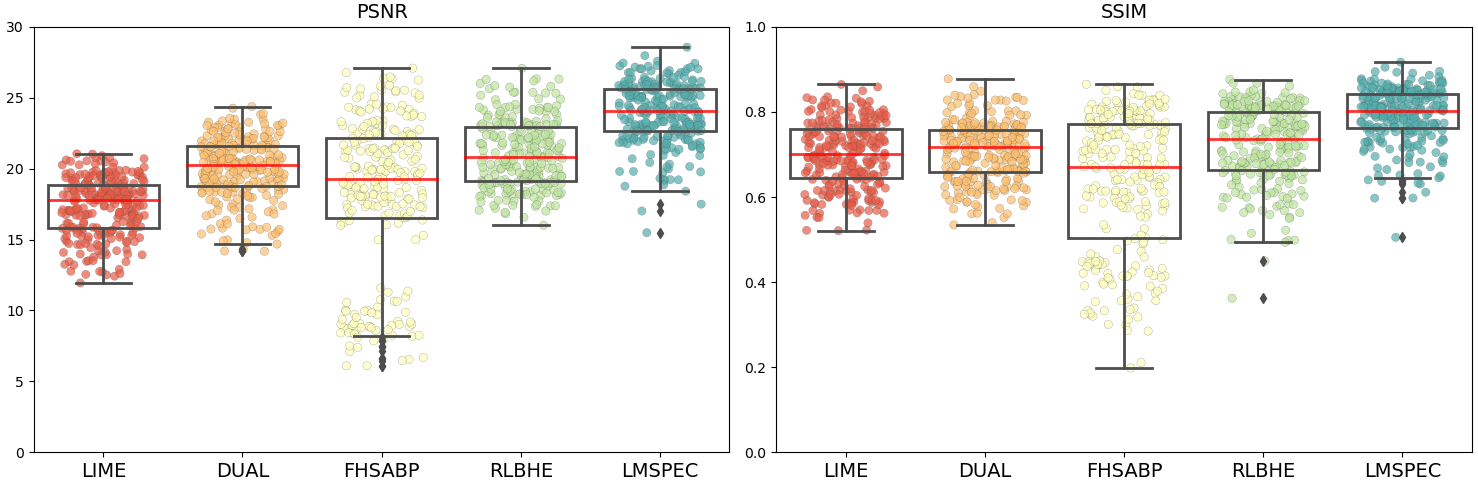}
         \caption{Underexposure.}
         \label{underexp}
     \end{subfigure}
        \caption{Reference-based PSNR and SSIM evaluation results.}
    \label{fig:boxplots}
\end{figure}

Table \ref{tab:t2} reports the quantitative results of each method applied to our dataset, in which numbers in bold were the best results for each metric. As it can be inferred from the table, LMSPEC achieves a very competitive performance for both types of exposure errors in terms of of noise removal and the preservation of the structure of the image.

\begin{figure}[!t]
    \centering
     \includegraphics[scale=0.45]{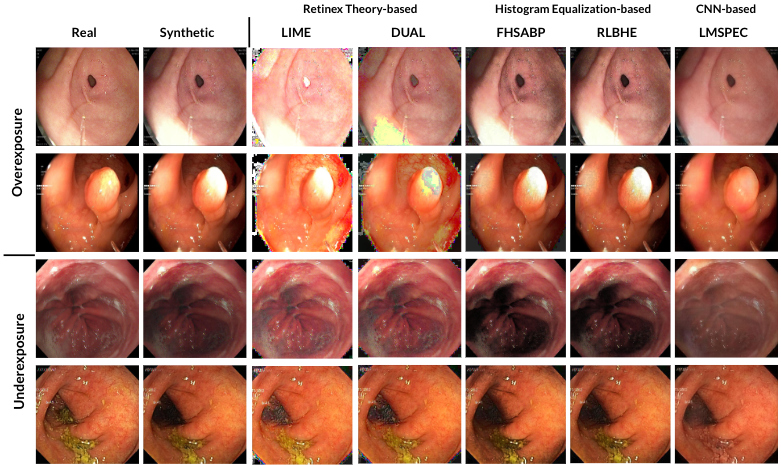}
    \caption{Results of our methods: first column real images, second column GANs-made synthetic images with exposure error, and third column the image corrected. First pair of rows, we show two overexposed cases, and second pair underexposed cases.}
    \label{fig:LMSPEC}
\end{figure}

On the other hand, Figure \ref{fig:boxplots} shows distribution and density of the computed metrics (in the form of box and plots) obtained for every applied enhanced method.  It turns out that LMSPEC got higher values for both compared metrics, and the results tend to be more robust as they show very low variance. Figure \ref{fig:LMSPEC} shows qualitatively results of enhancement with all methods. 

This qualitative comparison is very much required as by comparing methods in terms of MSE and SSIM might suggest that the the Retinex Theory-based DUAL method perform as well as LMSPEC. However, as it can be observed in Figure \ref{fig:LMSPEC}, this is far from the truth, as a comparison between the 4th and 7th columns shows: DUAL is in general not able to remove overexposure artefacts, and in some cases it makes the problem worse. 

\vspace{-2mm}

Another advantage of LMSPEC over DUAL and other more traditional methods pertains the processing or inference time. As it can be observed in Table \ref{tab:t2}, more traditional methods require very long processing times (44 and 30 seconds in average for image for LIME and DUAL, respectively) whereas LMSPEC requires 0.13 seconds, attaining a very good performance (around $7.6 fps$). 

\vspace{-4mm}

\section{Conclusion and Future Work}
\label{Conclusion_FW}

In this paper we have proposed a pipeline for generating a novel dataset which aim is to provide a baseline for comparing image enhancement methods. The dataset was built leveraging images frm other publicly available datasets for finding suitable uncorrupted and under and over exposed images using an object detector. This intermediate dataset in the used for transferring the desired artifacts into the uncorrupted images, effectively enabing us to create a paired dataset that can be used for reference-based testing purposes. After a filtering process,  this dataset was validated by expert endoscopists in our team.

However, some of the synthetic images still present drawbacks such as noise, color changes and other types distortions. We consider that a longer training, in tandem with a larger dataset or data augmentation techniques would alleviate this issue. Nonetheless, we demonstrate that this dataset is versatile enough for testing traditional and more advanced image enhancement methods such as LMSPEC. 

As our results suggest, this method has shown remarkable results, but further tests and improvements are necessary to make it useful for endoscopic. For instance, the model can introduce some artifacts in some images; this aspect needs to be properly characterized and taken into account for designing novel loss functions (i.e. perceptual loss) that preserve color and texture more effectively. Secondly, there are some areas of improvement in order to make this model to be able to run in real time (i.e., 24 FPS).

\vspace{-0.3cm}
\section*{Acknowledgments}
\vspace{-0.1cm}
The authors wish to thank the AI Hub and the CIIOT at ITESM for their support for carrying the experiments reported in this paper in their NVIDIA's DGX computer.


%
%
%

\bibliographystyle{splncs04}
\bibliography{references}

\end{document}